\documentclass[aoas,MSNbibl,nameyear,dvips]{arximspdf}
\usepackage{dcolumn}
\usepackage{graphicx}

\doi{10.1214/11-AOAS494}
\volume{5}
\issue{4}
\pubyear{2011}
\firstpage{2630}
\lastpage{2650}

\makeatletter

\newcolumntype{d}[1]{D{.}{.}{#1}}

\newcommand{\tvb}{|\!|\!|}

\newcommand{\mBernoulli}{\operatorname{Bernoulli}}

\newcommand{\bX}{\mathbf X}
\newcommand{\by}{\mathbf y}
\newcommand{\bC}{\mathbf C}
\newcommand{\bS}{\mathbf S}
\newcommand{\bW}{\mathbf W}

\makeatother

\begin{document}
\begin{frontmatter}

\title{A sparse conditional Gaussian graphical model for analysis of
genetical genomics data\thanksref{T1}}
\runtitle{A sparse conditional Gaussian graphical model}

\thankstext{T1}{Supported in part by NIH R01ES009911 and R01CA127334.}

\begin{aug}
\author[A]{\fnms{Jianxin} \snm{Yin}\ead[label=e1]{yinj@mail.med.upenn.edu}}
\and
\author[A]{\fnms{Hongzhe} \snm{Li}\corref{}\ead[label=e3]{hongzhe@upenn.edu}}
\runauthor{J. Yin and H. Li}
\affiliation{University of Pennsylvania School of Medicine}
\address[A]{Department of Biostatistics\\
\quad and Epidemiology\\
University of Pennsylvania School\\
\quad of Medicine\\
Philadelphia, Pennsylvania 19104\\
USA\\
\printead{e1}\\
\hphantom{E-mail: }\printead*{e3}} 
\end{aug}

\received{\smonth{3} \syear{2010}}
\revised{\smonth{6} \syear{2011}}

%
\begin{abstract}
Genetical genomics experiments have now been routinely conducted to
measure both the genetic markers and gene expression data on the same
subjects. The gene expression levels are often treated as quantitative
traits and are subject to standard genetic analysis in order to
identify the gene expression quantitative loci (eQTL). However, the
genetic architecture for many gene expressions may be complex, and
poorly estimated genetic architecture may compromise the inferences of
the dependency structures of the genes at the transcriptional level. In
this paper we introduce a sparse conditional Gaussian graphical model
for studying the conditional independent relationships among a set of
gene expressions adjusting for possible genetic effects where the gene
expressions are modeled with seemingly unrelated regressions. We
present an efficient coordinate descent algorithm to obtain the
penalized estimation of both the regression coefficients and the
sparse concentration matrix. The corresponding graph can be used to
determine the conditional independence among a group of genes while
adjusting for shared genetic effects. Simulation experiments and
asymptotic convergence rates and sparsistency are used to justify our
proposed methods. By sparsistency, we mean the property that all
parameters that are zero are actually estimated as zero with
probability tending to one. We apply our methods to the analysis of a
yeast eQTL data set and demonstrate that the conditional Gaussian
graphical model leads to a more interpretable gene network than a
standard Gaussian graphical model based on gene expression data alone.
\end{abstract}

%
\begin{keyword}
\kwd{eQTL}
\kwd{Gaussian graphical model}
\kwd{regularization}
\kwd{genetic networks}
\kwd{seemingly unrelated regression}.
\end{keyword}

\end{frontmatter}

\section{Introduction}
Genetical genomics experiments have now been routinely conducted to
measure both the genetic variants and the gene expression data on the
same subjects. Such data have provided important insights into gene
expression regulations in both model organisms and humans
[\citet{BreKru05}, \citet{Schetal03}, \citet{CheSpi02}].
Gene expression levels are treated as quantitative traits and are
subject to standard genetic analysis in order to identify the gene
expression quantitative loci (eQTL). However, the genetic architecture
for many gene expressions may be complex due to possible multiple
genetic effects and gene--gene interactions, and poorly estimated
genetic architecture may compromise the inferences of the dependency
structures of genes at the transcriptional level
[\citet{Netetal10}]. For a given gene, typical analysis of such
eQTL data is to identify the genetic loci or single nucleotide
polymorphisms (SNPs) that are linked or associated with the expression
level of this gene. Depending on the locations of the eQTLs or the
SNPs, they are often classified as distal \textit{trans}-linked loci or
proximal \textit{cis}-linked loci [\citet{KenWan03},
\citet{Kenetal06}]. Although such a single gene analysis can be
effective in identifying the associated genetic variants, gene
expressions of many genes are in fact highly correlated due to either
shared genetic variants or other unmeasured common regulators. One
important biological problem is to study the conditional independence
among these genes at the expression level.

eQTL data provide important information about gene regulation and have
been employed to infer regulatory relationships among genes
[\citet{Zhuetal04}, \citet{BinHoe05},
\citet{CheEmmSto07}]. Gene expression data have been used for
inferring the genetic regulatory networks, for example, in the
framework of Gaussian graphical models (\textit{GGM})
[\citet{SchStr05}, \citet{Segetal05}, \citet{LiGui06},
\citet{PenZhoZhu09}]. Graphical models use graphs to represent
dependencies among stochastic variables. In particular, the
\textit{GGM} assumes that the multivariate vector follows a
multivariate normal distribution with a~particular structure of the
inverse of the covariance matrix, called the concentration matrix. For
such Gaussian graphical models, it is usually assumed that the patterns
of variation in expression for a given gene can be predicted by those
of a small subset of other genes. This assumption leads to sparsity
(i.e., many zeros) in the concentration matrix and reduces the problem
to well-known neighborhood selection or covariance selection problems
[\citet{Dem72}, \citet{MeiBuh06}]. In such a concentration
graph modeling framework, the key idea is to use partial correlation as
a~measure of the independence of any two genes, rendering it
straightforward to distinguish direct from indirect interactions. Due
to high-dimensionality of the problem, regularization methods have been
developed to estimate the sparse concentration matrix where a sparsity
penalty function such as the~$L_1$ penalty or SCAD penalty is often
used on the concentration matrix [\citet{LiGui06},
\citet{FriHasTib08}, \citet{FanFenWu09}]. Among these
methods, the coordinate descent algorithm of \citet{FriHasTib08},
named \textit{glasso}, provides a computationally efficient method for
performing the Lasso-regularized estimation of the sparse concentration
matrix.

Although the standard \textit{GGM}s can be used to infer the
conditional dependency structures using gene expression data alone
from eQTL experiments, such models ignore the effects of genetic
variants on the means of the expressions, which can compromise the
estimate of the concentration matrix, leading to both false positive
and false negative identifications of the edges of the Gaussian
graphs. For example, if two genes are both regulated by the same
genetic variants, at the gene expression level, there should not be
any dependency of these two genes. However, without adjusting for
the genetic effects on gene expressions, a link between these two
genes is likely to be inferred. For eQTL data, we are interested in
identifying the conditional dependency among a set of genes after
removing the effects from shared regulations by the markers. Such a
graph can truly reflect gene regulation at the expression level.

In this paper we introduce a sparse conditional Gaussian
graphical model (\textit{cGGM}) that
simultaneously
identifies the genetic variants associated with gene expressions and
constructs a sparse Gaussian graphical model based on eQTL data.
Different from the standard \textit{GGM}s that assume constant means,
the \textit{cGGM} allows the means to depend on covariates or genetic
markers. We consider a set of regressions of gene expression in
which both regression coefficients and the error concentration
matrix have many zeros. Zeros in regression coefficients arise when
each gene expression only depends on a very small set of genetic
markers; zeros in the concentration matrix arise since the gene
regulatory network and therefore the corresponding concentration
matrix is sparse. This approach is similar in spirit to the
seemingly unrelated regression (SUR) model of \citet{Zel62} in
order to improve the estimation efficiency of the effects of genetic
variants on gene expression by considering the residual correlations
of the gene expression of many genes. In the analysis of eQTL data, we
expect sparseness in both the regression coefficients and also the
concentration matrix. We propose to develop a regularized
estimation procedure to simultaneously select the SNPs associated
with gene expression levels and to estimate the sparse concentration
matrix. Different from the original SUR model of \citet{Zel62} that
focuses on improving the estimation efficiency of the regression
coefficients, we focus more on estimating the sparse concentration
matrix adjusting for the effects of the SNPs on mean expression
levels. We develop an efficient coordinate descent algorithm to
obtain the penalized estimates and present the asymptotic results
to justify our estimates.

In the next sections we first present the formulation of the
\textit{cGGM} for both the mean gene expression levels and the concentration
matrix. We then present an efficient coordinate descent algorithm to
perform the regularized estimation of the regression coefficients
and concentration matrix. Simulation experiments and asymptotic
theory are used to justify our proposed methods. We apply the
methods to an analysis of a yeast eQTL data set. We conclude the
paper with a brief discussion. All the proofs are given in the
supplementary material [\citet{YinLi11}].

\section{The sparse \textit{cGGM} and penalized likelihood estimation}
\subsection{The sparse conditional Gaussian graphical model}
$\!\!$Suppose
we have~$n$
independent observations from a population of a vector $(\by^\prime,
\mathbf x^\prime)$, where $\by$ is a~$p\times1$ random vector of gene
expression levels of $p$ genes and $\mathbf x$ is a~$q\times1$ vector of
the numerically-coded SNP genotype data for $q$ SNPs. Furthermore,
suppose that conditioning on $\mathbf x$, $\by$ follows a multivariate
normal distribution,
%
%
\begin{equation}\label{SSUR}
\by|\mathbf x \sim\mathcal{N}(\Gamma\mathbf x, \Sigma),
\end{equation}
where $\Gamma$ is a $p\times
q$ coefficient matrix for the means and the covariance matrix
$\Sigma$ does not depend on $\mathbf x$. We are interested in both the
effects of the SNPs on gene expressions $\Gamma$ and the
conditional independence structure of~$\by$ adjusting for the
effects of $\mathbf x$, that is, the Gaussian graphical model for
$\by=(\by_1,\ldots, \by_p)$ conditional on $\mathbf x$. In applications
of gene expression data analysis, we are more interested in the
concentration matrix $\Theta=\Sigma^{-1}$ after their shared
genetic regulators are accounted for. It has a nice interpretation
in the Gaussian graphical model, as the $(i, j)$-element
is directly related to the partial
correlation between the $i$th and $j$th components of $\by$ after
their potential joint genetic regulators are adjusted. In the
Gaussian graphical model with undirected graph $(V,E)$, vertices $V$
correspond to components of the vector $\by$ and edges $E = \{e_{ij}
, 1 \le i,j \le p\}$ indicate the conditional dependence among
different components of $\by$. The edge $e_{ij}$ between $\by_i$ and
$\by_j$ exists if and only if $\theta_{ij} \ne0$, where
$\theta_{ij}$ is the $(i, j)$-element of $\Theta$. We emphasize that
in the graph representation of the random variable $\by$, the nodes
include only the genes and the markers are not part of the graph. We
call this the sparse conditional Gaussian graph model (\textit{cGGM})
of the genes.
Hence, of
particular interest is to identify zero entries in the concentration
matrix. Note that instead of assuming a constant mean as in the
standard \textit{GGM}, model (\ref{SSUR}) allows heterogeneous means.

In eQTL experiments, each row of $\Gamma$ and the concentration
matrix $\Theta$ are expected to be sparse and our goal is to
simultaneously learn the Gaussian graphical model as defined by the
$\Theta$ matrix and to identify the genetic variants associated with
gene expressions $\Gamma$ based on $n$ independent observations of
$(\by_i^\prime, \mathbf x_i^\prime), i=1,\ldots, n$. From now on,
we use
$\by_i$ to denote the vector of gene expression levels of the $p$
genes and $\mathbf x_i$ to denote the vector of the genotype codes of the
$q$ SNPs for the $i$th observation unless otherwise specified.
Finally, let $\bX=(\mathbf x_1^\prime,\ldots, \mathbf x_n^\prime)$
be the genotype
matrix and $\bar{\mathbf x}=1/n \sum_{i=1}^n \mathbf x_i$.

\subsection{Penalized likelihood estimation}
Suppose that we have $n$ independent observation $(\by_i^\prime,
\mathbf x_i^\prime)$ from the \textit{cGGM} (\ref{SSUR}). Let
$\bC_Y=1/n\sum_{i=1}^n \by_i\by_i^\prime$, $\bC_{YX}=1/n\times\sum_{i=1}^n
\by_i\mathbf x_i^\prime$ and
$\bC_X=1/n\sum_{i=1}^n \mathbf x_i\mathbf x_i^\prime$.
Then the negative of the logarithm of the likelihood function
corresponding to the \textit{cGGM} model can be written as
\[
l(\Xi)=-\log\det\Theta+
\operatorname{tr}\{\bC_Y \Theta-\bC_{YX} {\Gamma}^\prime\Theta-\Gamma
\bC_{YX}^\prime\Theta+\Gamma\bC_X{\Gamma}^\prime\Theta\},
\]
where $\Xi=(\Theta, \Gamma)$ represents the associated parameters in
the \textit{cGGM}.

The Hessian matrix of the negative log-likelihood
function $l(\Xi)$ is
\[
Hl(\Xi)=
\pmatrix{
\Theta^{-1}\otimes\Theta^{-1} & -2\bC_{YX}\otimes I_p + 2(\Gamma
\bC_X)\otimes I_p\cr
-2\bC_{YX}^\prime\otimes I_p + 2(\bC_X{\Gamma}^\prime)\otimes I_p &
2\bC_X\otimes\Theta}
\]
(see Proposition 1 in the
supplementary material [\citet{YinLi11}], Section~3). In
addition, $l(\Xi)$ is a bi-convex function of $\Gamma$ and $\Theta$.
In words, this means that for any fixed $\Theta$, $l(\Xi)$ is a
convex function of $\Gamma$, and for any $\Gamma$, $l(\Xi)$ is a~convex function of $\Theta$.
When $n> \max(p,q)$, the global minimizer of~$l(\Xi)$ is given
by
\[
\cases{
\tilde{\Theta}^{-1} = \bC_Y-\bC_{YX}\bC_X^{-1}\bC
_{YX}^\prime, \cr
\tilde{\Gamma} = \bC_{YX}\bC_X^{-1}.}
\]

Under the penalized likelihood framework, the estimate of the
$\Gamma$ and $\Sigma$ in model (\ref{SSUR}) is the solution to the
following optimization problem:
%
%
\begin{equation}\label{objadap}
\min\biggl\{\operatorname{pl}(\Xi) \equiv\,{-}\,\log
\det
\Theta\,{+}\,\operatorname{tr}(\bS_{\Gamma} \Theta)\,{+}\,\lambda\sum_{s,t} \operatorname{pen}_1(\gamma
_{st})\,{+}\,\rho
\sum_{t,t'}\operatorname{pen}_2(\theta_{tt'})\biggr\},\hspace*{-30pt}
\end{equation}
where $\operatorname{pen}_1(\cdot)$ and
$\operatorname{pen}_2(\cdot)$ denote the generic penalty functions, $\gamma_{st}$ is
the $st$th element of the $\Gamma$ matrix and $\theta_{tt'}$ is the
$tt'$th element of the $\Theta$ matrix, and
\begin{eqnarray*}
\mathbf{S}_{\Gamma} &=& \frac{1}{n}\sum_{i=1}^n(\by_i-\Gamma
\mathbf x_i)(\by_i-\Gamma\mathbf x_i)^\prime\\
&=& \bC_Y-\bC_{YX}{\Gamma}^\prime-\Gamma\bC_{YX}^\prime+\Gamma
\bC_X{\Gamma}^\prime.
\end{eqnarray*}
Here $\rho$ and $\lambda$ are the two
tuning parameters that control the sparsity of the sparse \textit
{cGGM}. We consider in this paper both the Lasso or $L_1$ penalty
function $\operatorname{pen}(x)=|x|$ [\citet{Tib96}] and the adaptive Lasso
penalty function $\operatorname{pen}(x)=|x|/|\tilde{x}|^{\gamma}$ for some
$\gamma>0$ and any consistent estimate of $x$, denoted by
$\tilde{x}$ [\citet{Zou06}]. In this paper we use $\gamma=0.5$.

\subsection{An efficient coordinate descent algorithm for the sparse
\textit{cGGM}}
We present an algorithm for the optimization problem (\ref{objadap})
with Lasso penalty function for $\operatorname{pen}_1(\cdot)$ and $\operatorname{pen}_2(\cdot)$. A
similar algorithm can be developed for the adaptive Lasso penalty
with simple modifications. Under\vadjust{\goodbreak} this penalty function, the
objective function is then
%
%
\begin{equation}\label{loglik}
\max\{\log
\det\Theta- \operatorname{tr}(\bS_{\Gamma}
\Theta)-\lambda\|\Gamma\|_1-\rho\|\Theta\|_1\}.
\end{equation}
The
subgradient equation for maximization of the log-likelihood
(\ref{loglik}) with respect to $\Theta$ is
%
%
\begin{equation}
\Theta^{-1}-\bS_{\Gamma}-\rho\Lambda= 0,
\end{equation}
where
$\Lambda_{ij}\in \operatorname{sgn}(\Theta_{ij})$. If $\Gamma$ is known,
\citet{BanElGdAs08} and \citet{FriHasTib08} have cast the
optimization problem (\ref{loglik}) as a block-wise coordinate
descent algorithm, which can be formulated as $p$ iterative Lasso
problems. Before we proceed, we first introduce some notation to
better represent the algorithm. Let $\bW$ be the estimate of
$\Sigma$. We partition $\bW$ and $\bS_{\Gamma}$ as
\[
\bW=\pmatrix{
\bW_{11} & w_{12}\vspace*{2pt}\cr
w_{12}^\top& w_{22}},\qquad
\bS_{\Gamma}=
\pmatrix{
\bS_{11} & s_{12}\vspace*{2pt}\cr s_{12}^\top
& s_{22}}.
\]
\citet{BanElGdAs08} show that the solution for $w_{12}$
satisfies
\[
w_{12}=\mathop{\arg\min}_y (y^T W_{11}^{-1}y\dvtx
\|y-s_{12}\|_{\infty} \le\rho),
\]
which by convex duality is
equivalent to solving the dual problem
%
%
\begin{equation}\label{glasso_step}
\hat\beta=\mathop{\arg\min}_{\beta}\biggl(\frac{1}{2}
\|\bW_{11}^{1/2}\beta-b\| ^2 + \rho\|\beta\|_1 \biggr),
\end{equation}
where $b=\bW_{11}^{-1/2}s_{12}$. Then
the solution for $w_{12}$ can be obtained via the solution of the
Lasso problem
and
through the relation $w_{12}=\bW_{11}\beta$. The estimate for
$\Theta$ can also be updated in this block-wise manner
very efficiently through the relationship $\bW\Theta=\mathbf{I}$
[\citet{FriHasTib08}].

After we finish an updating cycle for $\Theta$, we can proceed to
update the estimate of $\Gamma$. Since the object function of our
penalized log-likelihood is quadratic in $\Gamma$ given $\Theta$,
we can use a direct coordinate descent algorithm to get the penalized
estimate of $\Gamma$. For the ($i$, $j$)th entry of $\Gamma$,
$\gamma_{ij}$, note that for an arbitrary $q\times p$ matrix
$\mathbf{A}$, $\partial\operatorname{tr}(\Gamma A)/\partial
\gamma_{ij}=a_{ji}=e_j^\prime\mathbf{A} e_i$, where $e_j$ and $e_i$
are the corresponding base vector with $q$ and $p$ dimensions. So
the derivative of the penalized log-likelihood function
(\ref{loglik})with respect to $\gamma_{ij}$ is
%
%
\begin{equation}\label{eq_f}
2e_j^\prime(\bC_X {\Gamma}^\prime\Theta)e_i+\lambda
\operatorname{sgn}(\gamma_{ij})-2e_j^\prime(\bC_{YX}^\prime\Theta)e_i,
\end{equation}
where function $\operatorname{sgn}$ is defined as
\[
\operatorname{sgn}(t)=\cases{
1, &\quad if $t>0$,\cr
0, &\quad if $t=0$, and \cr
-1, &\quad if $t<0$.}
\]

Setting   equation (\ref{eq_f}) to zero, we get the updating
formula for $\gamma_{ij}$:
%
%
\begin{equation}\label{update_f}
\hat\gamma_{ij}=\operatorname{sgn}(g_{ij})\frac{(|g_{ij}|-\lambda
)_+}{2(e_j^\prime
\bC_X e_j)(e_i^\prime\Theta e_i)},
\end{equation}
where\vspace*{1pt}
$g_{ij}=2\{e_j^\prime(\bC_{YX}^\prime\Theta)e_i+(e_j^\prime
\bC_X e_j)(e_i^\prime\Theta e_i)\tilde{\gamma}_{ij}-e_j^\prime
(\bC_X \tilde{\Gamma}^\prime\Theta)e_i\}$ and $\tilde{\Gamma}$,
$\tilde{\gamma}_{ij}$ are the estimates in the last step of the
iteration.

Taking these two updating steps together, we have the following
coordinate descent-based regularization algorithm to fit the sparse
\textit{cGGM}:\vspace*{8pt}

\textit{The Coordinate Descent Algorithm for the sparse \textit{cGGM}.}
\begin{longlist}[(1)]
\item[(1)] Start with $\Gamma=\bC_{YX}\bC_X^{-1}$ and
$\bW=\bC_Y-\bC_{YX}\bC_X^{-1}\bC_{YX}^\prime+\rho I$. If $\bC_X$ is
not invertible, use $\Gamma=0$ and $\bW=\bC_Y+\rho I$ instead.
\item[(2)] For each $j=1, 2,\ldots, p$, solve the Lasso problem
(\ref{glasso_step}) under the current estimate of $\Gamma$. Fill in
the corresponding row and column of $\bW$
using $w_{12}=\bW_{11}\hat\beta$. Update $\hat\Theta$.
\item[(3)] For each $i=1, 2,\ldots, p$, and $j=1, 2,\ldots, q$
update each entry
$\hat\gamma_{ij}$ in $\hat\Gamma$ using the formula
(\ref{update_f}), under the current estimate for $\Theta$.
\item[(4)] Repeat step (2) and step (3) until convergence.\vspace*{1pt}
\item[(5)] Output the estimate $\hat\Theta$, $\hat\bW$ and $\hat
\Gamma$.
\end{longlist}

The adaptive version of the algorithm can be derived in the same
steps with adaptive penalty parameters and is omitted here. Note
that when $\Gamma=0$, this algorithm simply reduces to the \textit
{glasso} or the adaptive \textit{glasso} (\textit{aglasso}) algorithm of
\citet{FriHasTib08}. A similar algorithm was used in
\citet{RotLevZhu10} for sparse multivariate regressions.
Proposition 2 in the
supplementary material [\citet{YinLi11}] proves that the above
iterative algorithm for minimizing $\operatorname{pl}(\Xi)$ with respective to
$\Gamma$ and $\Theta$ converges to a stationary point of $\operatorname{pl}(\Xi)$.

While the iterative algorithm reaches a stationary point of
$\operatorname{pl}(\Xi)$, it is not guaranteed to reach the global minimum. Since
the objective function of the optimization problem (\ref{objadap})
is not always convex in $(\Gamma,\Theta)$, it is convex in either
$\Gamma$ or $\Theta$ with the other fixed. There are potentially
many stationary points due to the high-dimensional nature of the
parameter space. We also note a few straightforward properties of
the iterative procedure, namely, that each iteration monotonically
decreases the penalized negative log-likelihood and the order of
minimization is unimportant. Finally, the computational complexity
of this algorithm is $O(pq)$ plus the complexity of the \textit{glasso}.

\subsection{Tuning parameter selection}

The tuning parameters $\rho$ and $\lambda$ in the penalized
likelihood formulation (\ref{objadap}) determine the sparsity of the
\textit{cGGM} and have to be tuned. Since we focus on estimating the
sparse precision matrix and the sparse regression coefficients, we
use the Bayesian information criterion (BIC) to choose these two
parameters. The BIC is defined as
%
\[
\operatorname{BIC}(\hat\Theta,\hat\Gamma)=-n\log(|\hat
\Theta|)+n\operatorname{tr}(\hat\Theta
S_{\hat\Gamma})+\log(n)(s_n/2+p_n+k_n),
\]
where $p_n$ is the dimension of $\by$, $s_n$ is the number of
nonzero off-diagonal elements of $\hat\Theta$ and $k_n$ is the number of
nonzero elements of $\hat\Gamma$. The BIC has been shown to
perform well for selecting the tuning parameter of the penalized
likelihood estimator [\citet{WanLiTsa07}] and has been applied
for tuning parameter selection for \textit{GGMs} [\citet{PenZhoZhu09}].


\section{Theoretical properties}
Sections 4 and 5 in the
supplementary material [\citet{YinLi11}] state and prove
theoretical properties of the proposed penalized estimates of the
sparse \textit{cGGM}: its asymptotic distribution, the oracle
properties when $p$ and $q$ are fixed as $n\rightarrow\infty$ and
the convergence rates and sparsistency of the estimators when
$p=p_n$ and $q=q_n$ diverge as $n\rightarrow\infty$. By
sparsistency, we mean the property that all parameters that are zero
are actually estimated as zero with probability tending to one [\citet{LamFan09}].

We observe that the asymptotic bias for $\hat\Theta$ is at the same
rate as \citet{LamFan09} for sparse \textit{GGM}s, which is
$(p_n+s_n)/n$ multiplied by a~logarithm factor $\log p_n$, and goes
to zero as long as $(p_n+s_n)/n$ is at a~rate of $O\{(\log
p_n)^{-k}\}$ with some $k>1$. The total square errors for
$\hat\Gamma$ are at least of rate $k_n/n$ since each of the $k_n$
nonzero elements can be estimated with rate $n^{-1/2}$. The price we
pay for high-dimensionality is a logarithmic factor $\log(p_nq_n)$.
The estimate $\hat\Gamma$ is consistent as long as $k_n/n$ is at a
rate of $O\{(\log p_n+\log q_n)^{-l}\}$ with some $l>1$.

\section{Monte Carlo simulations}
In this section we present results from Monte Carlo simulations to
examine the performance of the proposed estimates and to compare it
with the \textit{glasso} procedure for estimating the Gaussian
graphical models using only the gene expression data. We also
compare the \textit{cGGM} with a modified version of the neighborhood
selection procedure of \citet{MeiBuh06},
where each gene is regressed on other genes and also the genetic
markers using the Lasso regression, and a link is defined between
gene $i$ and $j$ if gene $i$ is selected for gene $j$ and gene $j$
is also selected by gene $i$. We call this procedure the multiple
Lasso (\textit{mLasso}). Note that the \textit{mLasso} does not provide
an estimate of the concentration matrix. For adaptive procedures,
the MLEs of both the regression coefficients and the concentration
matrix were used for the weights when $p<n$ and $q<n$. For each
simulated data set, we chose the tuning parameters $\rho$ and
$\lambda$ based on the BIC.

To
compare the performance of different estimators for the
concentration matrix, we used the quadratic loss function
\[
\mathrm{LOSS}(\Theta, \hat\Theta)=\operatorname{tr}(\Theta^{-1}\hat
\Theta-I)^2,
\]
where $\hat\Theta$ is an estimate of the true concentration matrix
$\Theta$. We also compared $\|\Delta\|_\infty$,
$\tvb\Delta\tvb_\infty$, $\|\Delta\|$ and $\|\Delta\|_F$, where
$\Delta=\Theta-\hat\Theta$ is the difference between the true
concentration matrix and its estimate,
$\|A\|=\max\{\|Ax\|/\|x\|, x \in R^p, x\ne0\}$ is the
operator or spectral norm of a matrix $A$, $\|A\|_{\infty}$ is the
element-wise $l_{\infty}$ norm of a matrix $A$, $\tvb
A\tvb_{\infty}=\max_{1\leq i\leq p} \sum_{j=1}^q |a_{ij}|$
for $A=(a_{ij})_{p\times q}$ is\vspace*{2pt} the matrix $l_{\infty}$ norm of a
matrix $A$, and $\|A\|_F$ is the Frobenius norm, which is the
square-root of the sum of the squares of the entries of~$A$. In
order to compare how different methods recover the true graphical
structures, we considered the Hamming distance between the estimated
and the true concentration matrix, defined as $\operatorname
{DIST}(\Theta,
\hat\Theta)=\sum_{i, j} |I(\theta_{ij}\neq0)-\allowbreak I(\hat\theta
_{ij}\neq
0)|$, where $\theta_{ij}$ is the $(i, j)$th entry of $\Theta$ and
$I(\cdot)$ is the indicator function. Finally, we considered the
specificity (SPE), sensitivity(SEN) and Matthews correlation
coefficient (MCC) scores, which are defined as follows:
\begin{eqnarray*}
\mathrm{SPE}&=&\frac{\mathrm{TN}}{\mathrm{TN}+\mathrm{FP}},\qquad
\mathrm{SEN}=\frac{\mathrm{TP}}{\mathrm{TP}+\mathrm{FN}},
\\
\mathrm{MCC}&=&\frac{\mathrm{TP}\times \mathrm{TN}- \mathrm{FP}\times
\mathrm{FN}}{\sqrt{(\mathrm{TP}+\mathrm{FP})(\mathrm{TP}
+\mathrm{FN})(\mathrm{TN}+\mathrm{FP})(\mathrm{TN}+\mathrm{FN})}},
\end{eqnarray*}
where TP, TN, FP and FN are the numbers of true positives, true
negatives, false positives and false negatives in identifying the
nonzero elements in the concentration matrix. Here we consider the
nonzero entry in a sparse concentration matrix as ``positive.''

\subsection{Models for concentration matrix and generation of data}
In the following simulations, we considered a general sparse
concentration matrix, where we randomly generated a link (i.e.,
nonzero elements in the concentration matrix, indicated by
$\delta_{ij}$) between variables $i$ and $j$ with a success
probability proportional to $1/p$. Similar to the simulation setup
of \citet{LiGui06}, \citet{FanFenWu09} and \citet{PenZhoZhu09}, for each link, the corresponding entry in the concentration
matrix is generated uniformly over $[-1, -0.5]\cup[0.5, 1]$. Then
for each row, every entry except the diagonal one is divided by the
sum of the absolute value of the off-diagonal entries multiplied by
1.5. Finally, the matrix is symmetrized and the diagonal entries are
fixed at 1. To generate the $p\times q$ coefficient matrix
$\Gamma=(\gamma_{ij})$, we first generated a $p\times q$ sparse
indicator matrix $\Delta=(\delta_{ij})$, where $\delta_{ij}=1$ with
a probability proportional to $1/q$. If $\delta_{ij}=1$, we
generated $\gamma_{ij}$ from $\operatorname{Unif}([v_m, 1]\cup[-1, -v_m])$,
where $v_m$ is the minimum absolute nonzero value of $\Theta$
generated.

After $\Gamma$ and $\Theta$ were generated, we generated the marker
genotypes $X=(X_1,\ldots, X_q)$ by assuming $X_i\sim\mBernoulli(1,
\frac{1}{2}),\mbox{ for }i=1,\ldots, q$. Finally, gi\-ven~$\mathbf
x$, we
generated $\by$ the multivariate normal distribution $Y|X \sim
\mathcal{N}(\Gamma X, \Sigma)$. For a given model and a given
simulation, we generated a data set of $n$ \textit{i.i.d.} random vectors
$(X, Y)$. The simulations were repeated 50 times.

%
\begin{table}
\caption{Comparison of the performances of the \textit{cGGM}, adaptive
\textit{cGGM} (a\textit{cGGM}), graphical Lasso (\textit{glasso}),
adaptive graphical Lasso (a\textit{glasso}) and a modified neighborhood
selection procedure using multiple Lasso (\textit{mLasso}) for models
1--3 when $p<n$ based on 50 replications, where $n$ is the sample size,
$p$ is the number of genes and $q$ is the number of markers. For each
measurement, mean is given based on 50 replications. Simulation
standard errors are given in the supplementary material [Yin and Li
(\protect\citeyear{YinLi11})]}\label{simu.tb2}
\begin{tabular*}{\tablewidth}{@{\extracolsep{\fill}}ld{2.2}cccd{1.2}d{3.2}ccc@{}}
\hline
\multicolumn{1}{c}{} & \multicolumn{5}{c}{\textbf{Estimation of}
$\bolds{\Theta}$} &
\multicolumn{4}{c@{}}{\textbf{Graph selection}}\\[-4pt]
& \multicolumn{5}{c}{\hrulefill} & \multicolumn{4}{c@{}}{\hrulefill}\\
\textbf{Method} & \multicolumn{1}{c}{\textbf{LOSS}}
& \multicolumn{1}{c}{$\bolds{\|\Delta\|_\infty}$} &
\multicolumn{1}{c}{$\bolds{\tvb\Delta\tvb_\infty}$} &
\multicolumn{1}{c}{$\bolds{\|\Delta\|}$} &
\multicolumn{1}{c}{$\bolds{\|\Delta\|_F}$} &
\multicolumn{1}{c}{\textbf{DIST}} & \multicolumn{1}{c}{\textbf{SPE}}
& \multicolumn{1}{c}{\textbf{SEN}} & \multicolumn{1}{c@{}}{\textbf{MCC}} \\
\hline
& \multicolumn{9}{c@{}}{Model 1: $(p,q,n)=(100,100,250)$, $\operatorname{pr}(\theta
_{ij}\ne0)=2/p$,
$\operatorname{pr}(\Gamma_{ij} \ne0)=3/q$}\\
[4pt]
\textit{cGGM} & 10.73 & 0.33 & 1.17 & 0.67 & 3.18 & 279.56 & 0.99 & 0.48
& 0.56 \\
\textit{acGGM} & 10.29 & 0.31 & 1.17 & 0.66 & 3.01 &313.48 & 0.99 & 0.42
& 0.50\\
\textit{glasso} & 19.17 & 0.69 & 1.89 & 1.12 & 5.19 &596.12 & 0.97 &
0.24 & 0.21\\
\textit{aglasso} & 17.93 & 0.69 & 1.89 & 1.11 & 4.98 &541.32 & 0.97 &
0.32 & 0.28 \\
\textit{mLasso} & \multicolumn{1}{c}{--} & \multicolumn{1}{c}{--} & \multicolumn{1}{c}{--} &\multicolumn{1}{c}{--} &\multicolumn{1}{c}{--} &309.50 & 0.99 & 0.38 & 0.48
\\[4pt]
& \multicolumn{9}{c@{}}{Model 2: $(p, q, n)=(50, 50, 250)$,
$\operatorname{pr}(\theta_{ij}\ne0)=2/p$, $\operatorname{pr}(\Gamma_{ij}
\ne0)=4/q$}\\
[4pt]
\textit{cGGM} & 5.15 & 0.37 & 1.30 & 0.72 & 2.36 & 106.88 & 0.98 &
0.69 &
0.66 \\
\textit{acGGM} & 4.62 & 0.29 & 1.14 & 0.63 & 1.97 & 83.20 & 0.99 &
0.66 &
0.71\\
\textit{glasso} & 13.95 & 0.75 & 2.12 & 1.20 & 4.57 & 391.84 & 0.87 &
0.37 & 0.18\\
\textit{aglasso} & 13.15 & 0.74 & 2.11 & 1.19 & 4.4 & 389.00 & 0.87 &
0.49 & 0.25 \\
\textit{mLasso} & \multicolumn{1}{c}{--} & \multicolumn{1}{c}{--} & \multicolumn{1}{c}{--} & \multicolumn{1}{c}{--} & \multicolumn{1}{c}{--} &185.68 & 0.95 & 0.60 & 0.48
\\[4pt]
& \multicolumn{9}{c@{}}{Model 3: $(p, q, n)=(25, 10, 250)$,
$\operatorname{pr}(\theta_{ij}\ne0)=2/p$,
$\operatorname{pr}(\Gamma_{ij} \ne0)=3.5/q$}\\
[4pt]
\textit{cGGM} & 1.70 & 0.24 & 0.90 & 0.52 & 1.21 & 67.08 & 0.91 & 0.76 &
0.62 \\
\textit{acGGM} & 1.58 & 0.22 & 0.87 & 0.49 & 1.12 & 56.36 & 0.94 &
0.72 &
0.65 \\
\textit{glasso} & 5.97 & 0.65 & 1.99 & 1.12 & 2.77 & 315.84 & 0.43 & 0.73
& 0.12 \\
\textit{aglasso} & 6.05 & 0.65 & 1.98 & 1.12 & 2.78 & 264.30 & 0.54 &
0.65 & 0.14\\
\textit{mLasso} & \multicolumn{1}{c}{--} & \multicolumn{1}{c}{--} &\multicolumn{1}{c}{--} & \multicolumn{1}{c}{--} & \multicolumn{1}{c}{--} & 111.28 & 0.84 & 0.68 & 0.44 \\
\hline
\end{tabular*}
\end{table}

\subsection{Simulation results when $p<n$ and $q<n$}
We first consider the setting when the sample size $n$ is larger
than the number of genes $p$ and the number of genetic markers $q$.
In particular, the
following three models were considered:

\begin{longlist}[Model 2:]
\item[Model 1:]
$(p,q,n)=(100, 100, 250)$, where $\operatorname{pr}(\theta_{ij}\ne
0)=2/p$,
$\operatorname{pr}(\Gamma_{ij} \ne0)=3/q$;

\item[Model 2:] $(p, q, n)=(50, 50, 250)$, where
$\operatorname{pr}(\theta_{ij}\ne0)=2/p$,
$\operatorname{pr}(\Gamma_{ij} \ne0)=4/q$;

\item[Model 3:] $(p, q, n)=(25, 10, 250)$, where
$\operatorname{pr}(\theta_{ij}\ne0)=2/p$,
$\operatorname{pr}(\Gamma_{ij} \ne0)=3.5/q$.
\end{longlist}

We present the simulation
results in Table \ref{simu.tb2}. Clearly, \textit{cGGM} provided
better estimates (in terms of the defined LOSS function and the four
metrics of ``closeness'' of the estimated and true matrices) of the
concentration matrix over \textit{glasso} for all three models considered
in all measurements. This is expected since \textit{glasso} assumes a
constant mean of the multivariate vector, which is not a
misspecified model. We also observed that the adaptive \textit{cGGM}
and adaptive \textit{glasso} both resulted in better estimates of the
concentration matrix, although the improvements were minimal. This
may be due to the fact that the MLEs of the concentration matrix
when $p$ is relatively large do not provide very informative weights
in the $L_1$ penalty functions.

In terms of graph structure selection, we first observed that
different values of the tuning parameter $\rho$ for the penalty on
the mean parameters resulted in different identifications of the
nonzero elements in the concentration matrix, indicating that the
regression parameters in the means indeed had effects on estimating
the concentration matrix. Table \ref{simu.tb2} shows that for all
three models, the \textit{cGGM} or the adaptive \textit{cGGM} resulted in
higher sensitivities, specificities and MCCs than the \textit{glasso}
or the adaptive \textit{glasso}. We observed that \textit{glasso} often
resulted in much denser graphs than the real graphs. This is
partially due to the fact that some of the links identified by \textit
{glasso} can be explained by shared common genetic variants. By
assuming constant means, in order to compensate for the model
misspecification, \textit{glasso} tends to identify many nonzero
elements in the concentration matrix and result in larger Hamming
distance between the estimate and the true concentration matrix.
The results indicate that by simultaneously considering the
effects of the covariates on the means, we can reduce both false
positives and false negatives in identifying the nonzero elements
of the concentration matrix.

The modified neighborhood selection procedure using multiple Lasso
accounts for the genetic effects in modeling the relationship among
the genes. It performed better than \textit{glasso} or adaptive \textit
{glasso} in graph structure selection, but worse than the \textit{cGGM}
or the adaptive \textit{cGGM}. This procedure, however, did not provide
an estimate of the concentration matrix.

\subsection{Simulation results when $p>n$}

In this section we consider the setting when $p>n$ and simulate
data from the following three models with values of $n$, $p$
and $q$ specified as follows:

\begin{longlist}[Model 4:]
\item[Model 4:]
$(p, q, n)\,{=}\,(1000, 200, 250)$,
$\operatorname{pr}(\Theta_{ij}\,{\ne}\,0)\,{=}\,1.5/p$,
$\operatorname{pr}(\Gamma_{ij}\,{\ne}\,0)\,{=}\,20/q$;

\item[Model 5:] $(p, q, n)\,{=}\,(800, 200, 250)$,
$\operatorname{pr}(\Theta_{ij}\,{\ne}\,0)\,{=}\,1.5/p$,
$\operatorname{pr}(\Gamma_{ij}\,{\ne}\,0)\,{=}\,25/q$;

\item[Model 6:] $(p, q, n)\,{=}\,(400, 200, 150)$,
$\operatorname{pr}(\Theta_{ij}\,{\ne}\,0)\,{=}\,2.5/p$,
$\operatorname{pr}(\Gamma_{ij}\,{\ne}\,0)\,{=}\,20/q$.
\end{longlist}

Note that for all three models, the graph structure is very sparse
due to the large number of genes considered.

Since in this setting we did not have consistent estimates of
$\Gamma$ or $\Omega$, we did not consider the adaptive \textit{cGGM} or
adaptive \textit{glasso} in our comparisons. Instead, we compared the
performance of \textit{cGGM}, \textit{glasso} and the modified
neighborhood selection procedure using multiple Lasso in terms of
estimation of the concentration matrix and graph structure
selection. The performances over 50 replications are reported in
Table \ref{simu.tbl} for the optimal tuning parameters chosen by the
%
%
\begin{table}
\caption{Comparison of the performances of the \textit{cGGM}, the
graphical Lasso (\textit{glasso}) and a modified neighbor selection
using multiple lasso (\textit{mLasso}) $\mbox{model } 4\sim \mbox{model } 6$ when $p>n$
based on 50 replications, where $n$ is the sample size, $p$ is the
number of genes and $q$ is the number of markers. For each measurement,
mean is given based on 50 replications. Simulation standard errors are
given in the supplementary material [Yin and Li
(\protect\citeyear{YinLi11})]}\label{simu.tbl}
\begin{tabular*}{\tablewidth}{@{\extracolsep{\fill}}ld{3.2}cccd{2.2}rccd{2.2}@{}}
\hline
\multicolumn{1}{c}{} & \multicolumn{5}{c}{\textbf{Estimation of}
$\bolds{\Theta}$} &
\multicolumn{4}{c@{}}{\textbf{Graph selection}}\\[-4pt]
& \multicolumn{5}{c}{\hrulefill} & \multicolumn{4}{c@{}}{\hrulefill}\\
\textbf{Method} & \multicolumn{1}{c}{\textbf{LOSS}}
& \multicolumn{1}{c}{$\bolds{\|\Delta\|_\infty}$} &
\multicolumn{1}{c}{$\bolds{\tvb\Delta\tvb_\infty}$} &
\multicolumn{1}{c}{$\bolds{\|\Delta\|}$} &
\multicolumn{1}{c}{$\bolds{\|\Delta\|_F}$} &
\multicolumn{1}{c}{\textbf{DIST}} & \multicolumn{1}{c}{\textbf{SPE}}
& \multicolumn{1}{c}{\textbf{SEN}} & \multicolumn{1}{c@{}}{\textbf{MCC}} \\
\hline
& \multicolumn{9}{c@{}}{Model 4: $(p, q, n)=(1000, 200,
250)$,
$\operatorname{pr}(\Theta_{ij}\ne0)=1.5/p$,
$\operatorname{pr}(\Gamma_{ij}\ne0)=20/q$}\\
[4pt]
\textit{cGGM} & 164.22 & 0.59 & 1.81 & 0.97 & 13.48
& 2\mbox{,}414.28 & 1.00 &
0.31 & 0.47 \\
\textit{glasso} & 257.12 & 0.71 & 2.86 & 1.31 & 19.82
& 23\mbox{,}746.98 &
0.98 & 0.08 & 0.02\\
\textit{mLasso}& \multicolumn{1}{c}{--} & \multicolumn{1}{c}{--}
& \multicolumn{1}{c}{--} &\multicolumn{1}{c}{--}
& \multicolumn{1}{c}{--}
&3\mbox{,}886.96 & 1.00 & 0.12 & 0.16 \\
[4pt]
& \multicolumn{9}{c@{}}{Model 5: $(p, q, n)=(800, 200, 250)$,
$\operatorname{pr}(\Theta_{ij}\ne0)=1.5/p$,
$\operatorname{pr}(\Gamma_{ij}\ne0)=25/q$}\\
[4pt]
\textit{cGGM} & 142.30 & 0.75 & 2.30 & 1.20 & 12.82 & 2\mbox{,}341.28
& 1.00 & 0.21 & 0.34 \\
\textit{glasso} & 219.33 & 0.76 & 2.97 & 1.40 & 18.39
& 20\mbox{,}871.44 & 0.97 & 0.07 & 0.02 \\
\textit{mLasso} & \multicolumn{1}{c}{--} &\multicolumn{1}{c}{--}
&\multicolumn{1}{c}{--} & \multicolumn{1}{c}{--} & \multicolumn{1}{c}{--}
& 23\mbox{,}750.04 & 0.96 & 0.61 & 0.19 \\
[4pt]
& \multicolumn{9}{c@{}}{Model 6: $(p, q, n)=(400, 200, 250)$,
$\operatorname{pr}(\Theta_{ij}\neq0)=2.5/p$,
$\operatorname{pr}(\Gamma_{ij}\neq0)=20/q$}\\
[4pt]
\textit{cGGM} & 48.73 & 0.44 & 1.55 & 0.77 & 6.86 & 2\mbox{,}044.52
& 1.00 & 0.05
& 0.21\\
\textit{glasso} & 87.32 & 0.69 & 2.72 & 1.22 & 11.01 & 9\mbox{,}258.92
& 0.95 &
0.03 & -0.01 \\
\textit{mLasso}
& \multicolumn{1}{c}{--} &\multicolumn{1}{c}{--}
&\multicolumn{1}{c}{--} &\multicolumn{1}{c}{--}
&\multicolumn{1}{c}{--} & 2\mbox{,}967.30 & 0.99 & 0.08 & 0.10\\
\hline
\end{tabular*}
\end{table}
BICs. For all three models, we observed much improved estimates of
the concentration matrix from the proposed \textit{cGGM} as reflected
by both smaller $L_2$ loss functions and different norms of the
difference between the true and estimated concentration matrices.
The \textit{mLasso} procedure did not provide estimates of the
concentration matrix.

In terms of graph structure selection, since \textit{glasso} does not
adjust for potential effects of genetic markers on gene expressions,
it resulted in many wrong identifications and much lower
sensitivities and smaller MCCs than the \textit{cGGM}. Compared to
the modified neighborhood selection using multiple Lasso, estimates
from the \textit{cGGM} have smaller Hamming distance and larger MCC
than \textit{mLasso}. In general, we observed that when $p$ is larger
than the sample size, the sensitivities from all three procedures
are much lower than the settings when the sample size is larger. For
models 5 and 6, \textit{mLasso} gave higher sensitivities but lower
specificities than \textit{cGGM} or \textit{glasso}. This indicates that
recovering the graph structure in a high-dimensional setting is
statistically difficult. However, the specificities are in general
very high, agreeing with our theoretical sparsistency result of the
estimates.

\section{Analysis of yeast eQTL data}
To demonstrate the proposed methods, we present results
from the analysis of a data set generated by \citet{BreKru05}. In this experiment, 112 yeast segregants, one from each
tetrad, were grown from a cross involving parental strains BY4716
and wild isolate RM11-1a. RNA was isolated and cDNA was hybridized
to microarrays in the presence of the same BY reference material.
Each array assayed 6,216 yeast genes. Genotyping was performed using
GeneChip Yeast Genome S98 microarrays on all 112 $F_1$ segregants.
These 112 segregants were individually genotyped at 2,956 marker
positions. Since many of these markers are in high linkage
disequilibrium, we combined the markers into 585 blocks where the
markers within a block differed by at most 1 sample. For each block,
we chose the marker that had the least number of missing values as
the representative marker.

Due to small sample size and limited perturbation to the biological
system, it is not possible to construct a gene network for all 6,216
genes. We instead focused our analysis on two sets of genes that are
biologically relevant: the first set of 54 genes that belong to the
yeast MAPK signaling pathway provided by the KEGG database
[\citet{Kanetal10}], another set of 1,207 genes of the
protein--protein interaction (PPI) network obtained from a previously
compiled set by \citet{Steetal02} combined with protein physical
interactions deposited in the Munich Information center for Protein
Sequences (MIPS). Since the available eQTL data are based on
observational data, given limited sample size and limited perturbation
to the cells from the genotypes, it is statistically not feasible to
learn directed graph structures among these genes. Instead, for each of
these two data sets, our goal is to construct a conditional independent
network among these genes at the expression levels based on the sparse
conditional Gaussian graphical model in order to remove the false links
by conditioning on the genetic marker information. Such graphs can be
interpreted as a projection of true signaling or a protein interaction
network into the gene space [\citet{BraFueMen02}, \citet{Kon}].

\subsection{Results from the \textit{cGGM}
analysis of 54 MAPK pathway genes}

\begin{figure}

\includegraphics{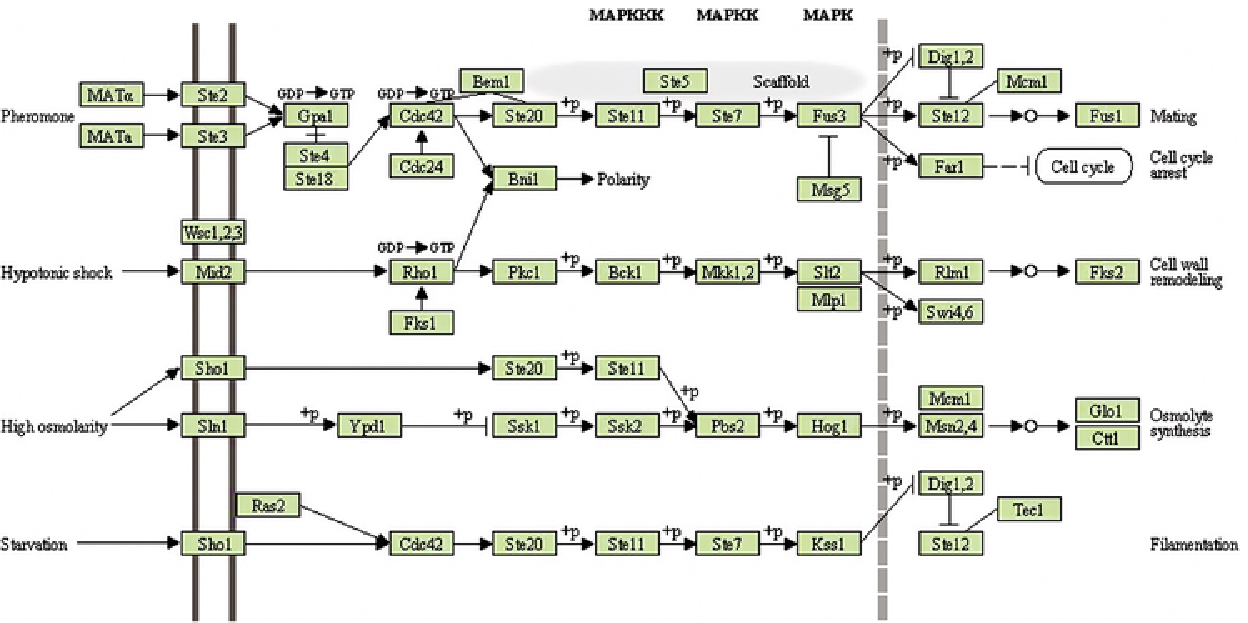}

\caption{The yeast MAPK pathway from the KEGG database
\protect\href{http://www.genome.jp/kegg/pathway/sce/sce04011.html}{http://www.genome.jp/}
\href{http://www.genome.jp/kegg/pathway/sce/sce04011.html}{kegg/pathway/sce/sce\textit{04011}.html}.}
\label{KEGG}
\end{figure}

The yeast genome encodes multiple MAP kinase orthologs, where Fus3
mediates cellular response to peptide pheromones, Kss1 permits
adjustment to nutrient-limiting conditions and Hog1 is necessary for
survival under hyperosmotic conditions. Last, Slt2/Mpk1 is
required for repair of injuries to the cell wall. A schematic plot
of this pathway is presented in Figure \ref{KEGG}. Note that this
graph only presents our current knowledge about the MAPK signaling
pathway. Since several genes such as Ste20, Ste12 and Ste7 appear at
multiple nodes, this graph cannot be treated as the ``true graph''
for evaluating or comparing different methods. In addition, although
some of the links are directed, this graph does not meet the
statistical definition of either a directed or undirected graph.
Rather than trying to recover the MAPK pathway structure, we chose
this set of 54 genes on the MAPK pathway to make sure that these
genes are potentially dependent at the expression level.

%

\begin{figure}
\begin{tabular}{@{}c@{}}

\includegraphics{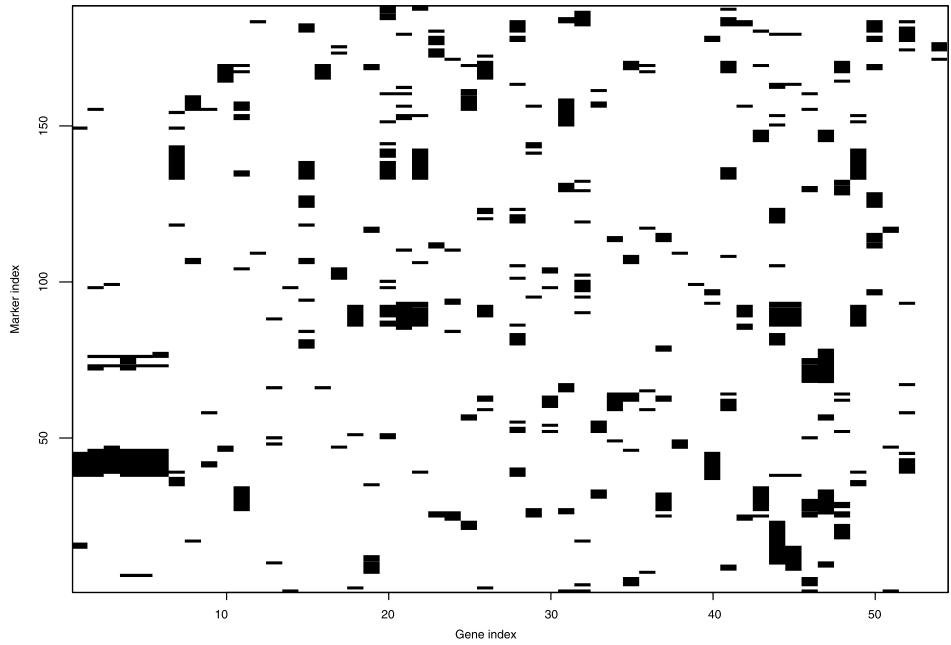}
\\
(a)\\[4pt]

\includegraphics{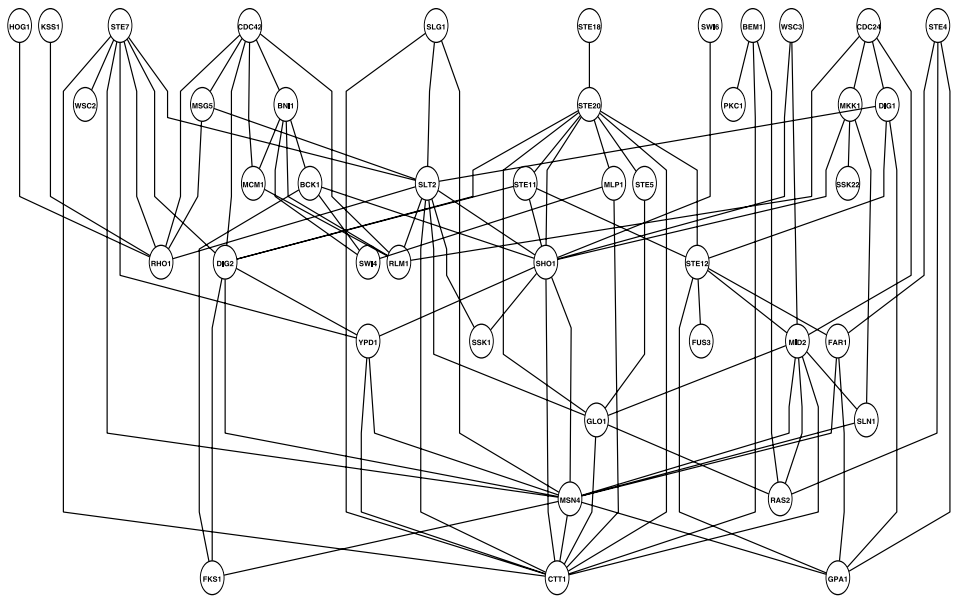}
\\
(b)
\end{tabular}
\caption{Analysis of yeast MAPK pathway. \textup{(a)} Association between
188 markers and 54 genes in the MAPK pathway based on simple
regression analysis. Black color indicates significant association
at $p$-$\mbox{value}<0.01$. \textup{(b)} The undirected graph of 43 genes
constructed based on the \textit{cGGM}.}
\label{MAPK}
\end{figure}

For each of the 54 genes, we first performed a linear regression
analysis for the gene expression level using each of the 585 markers
and selected those markers with a $p$-value of 0.01 or smaller. We
observed a total of 839 such associations between the 585 markers and
54 genes, indicating strong effects of genetic variants on expression
levels. We further selected 188 markers associated with the gene
expression levels of at least two out of the 54 genes, resulting in a
total of 702 such associations. In addition, many genes are associated
with multiple markers [see Figure \ref{MAPK}(a)]. This indicates that
many pairs of genes are regulated by some common genetic variants,
which, when not taken into account, can lead to false links of genes at
the expression level.

%

We applied our proposed \textit{cGGM} on this set of 54 genes and 188
markers and used the BIC to choose the tuning parameters. The BIC
selected $\lambda=0.28$ and $\rho=0.54$. With these tuning
parameters, the \textit{cGGM} procedure selected 188 nonzero elements
of the concentration matrix and therefore 94 links among these 54
genes. In addition, under the \textit{cGGM} model, 677 elements of the
regression coefficients $\Gamma$ are not zero, indicating the SNPs
have important effects on the gene expression levels of these genes.
The numbers of SNPs associated with the gene expressions range
from 0 to 17 with a mean number of 4. Figure \ref{MAPK}(b) shows
the undirected graph for 43 linked genes on the MAPK pathway based
on the estimated sparse concentration matrix from the \textit{cGGM}.
This undirected graph constructed based on the \textit{cGGM} can indeed
recover lots of links among the 54 genes on this pathway. For
example, the kinase Fus3 is linked to its downstream genes Dig1,
Ste12 and Fus1. The \textit{cGGM} model also recovered most of the
links to Ste20, including Bni1, Ste11, Ste12, Ste5 and Ste7. Ste20
is also linked to Cdc42 through Bni1. Clearly, most of the links in
the upper part of the MAPK signaling pathway were recovered by \textit
{cGGM}. This part of the pathway mediates cellular response to
peptide pheromones. Similarly, the kinase Slt2/Mpk1 is linked to its
downstream genes Swi4 and Rlm1. Three other genes on this second
layer of the pathway, Fks1, Rho1 and Bck1, are also closed linked.
These linked genes are related to cell response to hypotonic shock.

As a comparison, we applied the \textit{glasso} to the gene expression of
these 54 genes without adjusting the effects of genetic markers on
gene expressions and summarized the results in Table \ref{compare}.
The optimal tuning parameter $\lambda=0.145$ was selected based on
the BIC, which resulted in selection of 341 edges among the 54 genes
(i.e., 682 nonzero elements of the concentration matrix), including
all 94 links selected by the \textit{cGGM}. The difference of the
estimated graph structures between the \textit{cGGM} and \textit{glasso}
can be at least partially explained by the genetic variants
associated with the expression levels of multiple genes. Among these
247 edges that were identified by only the \textit{glasso}, 41 pairs of
genes were associated with at least one genetic variant. The \textit
{cGGM} adjusted the genetic effects on gene expression and therefore
did not identify these edges at the expression levels. Another
reason is that the \textit{glasso} assumes a constant mean vector for
gene expression, which clearly misspecified the model and led to the
selection of more links.

%
\begin{table}
\caption{Comparison of the links identified by the \textit{cGGM},
modified neighborhood selection using multiple Lasso (\textit{mLasso}),
the graphical Lasso (\textit{glasso}) for the genes of the
MAPK pathway and genes of the protein--protein interaction (PPI)
network. Shown in the table is the number of links that were
identified by the procedure indexed by row but were not identified
by the procedure indexed by column due to sharing of at least one
common genetic marker}\label{compare}
\begin{tabular*}{\tablewidth}{@{\extracolsep{4in minus 4in}}lcc@{}}
\hline
&
\includegraphics{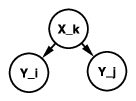}

&
\includegraphics{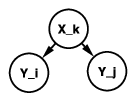}

\\
& \textbf{\textit{cGGM}} & \textbf{\textit{mLasso}} \\[-4pt]
& \multicolumn{2}{c@{}}{\hrulefill}\\
\multicolumn{1}{c}{}&\multicolumn{2}{c@{}}{\textbf{MAPK
pathway (PPI network)}}\\
\hline

\includegraphics{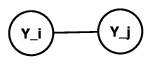}

\textit{cGGM} & -- & 0 (0)\hphantom{0}\\

\includegraphics{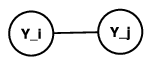}

\textit{mLasso} & 10 (218)\hphantom{0,}& --\\

\includegraphics{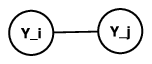}

\textit{glasso} & 41 (1,569) &
2 (66)\\
\hline
\end{tabular*}
\end{table}

We also compared the graph identified by the modified neighborhood
selection procedure of using multiple Lasso. Specifically, each
gene was regressed on all other genes and 188 markers using the
Lasso. Again, the BIC was used for selecting the tuning parameter.
This procedure identified a total of 45 links among the 54 genes. In
addition, a total of 33 associations between the SNPs and gene
expressions were identified. Among these 45 links, 36 were
identified by the \textit{cGGM} and 45 were identified by \textit{glasso}.

%
\begin{table}
\caption{Summary of degrees of the graphs constructed by three
different methods: \textit{cGGM}, the graphical Lasso (\textit{glasso})
and a modified neighborhood selection using multiple Lasso
(\textit{mLasso}), for the genes of the MAPK pathway and genes of the
protein--protein interaction (PPI) network}
\label{degree}
\begin{tabular*}{\tablewidth}{@{\extracolsep{\fill}}lcd{2.0}d{2.2}d{2.0}
ccd{2.2}d{2.0}@{}}
\hline\multicolumn{1}{c}{} & \multicolumn{4}{c}{\textbf{MAPK pathway}} &
\multicolumn{4}{c}{\textbf{PPI network}} \\[-4pt]
& \multicolumn{4}{c}{\hrulefill} & \multicolumn{4}{c@{}}{\hrulefill}\\
\textbf{Method} & \multicolumn{1}{c}{\textbf{Min}}
& \multicolumn{1}{c}{\textbf{Max}} & \multicolumn{1}{c}{\textbf{Mean}}
& \multicolumn{1}{c}{\textbf{Median}} & \multicolumn{1}{c}{\textbf{Min}}
& \multicolumn{1}{c}{\textbf{Max}} & \multicolumn{1}{c}{\textbf{Mean}}
& \multicolumn{1}{c@{}}{\textbf{Median}}\\
\hline
\textit{cGGM} & 0 & 11 & 3.48 & 3 & 0 & 57 & 19.94 & 21\\
glasso & 5 & 19& 12.63 & 13 & 5 & 60 & 31.46 & 32\\
\textit{mLasso} & 0 & 6& 1.67 & 1 & 0 & 12 & 3.18 & 3\\
\hline
\end{tabular*}
\end{table}

Table \ref{degree} shows a summary of the degrees of the graphs
estimated by these three procedures. It is clear that \textit{glasso}
resulted in a much denser graph than the neighborhood selection and
\textit{cGGM}, and the \textit{mLasso} tends to select few links.

\subsection{Results from the \textit{cGGM} analysis of 1,207 genes on yeast
PPI network}

We next applied the \textit{cGGM} to the yeast protein--protein
interaction network data obtained from a previously compiled set by
\citet{Steetal02} combined with protein physical interactions
deposited in MIPS. We further selected 1,207 genes with variance
greater than 0.05. Based on the most recent yeast protein--protein
interaction database BioGRID [\citet{Staetal11}], there are a
total of 7,619 links among these 1,207 genes. The BIC chose
$\lambda=0.34$ and $\rho=0.43$, which resulted in selection of 12,036
links out of a total of 727,821 possible links, which gives a sparsity
of 1.65\%. Results from comparisons with the two other procedures are
shown in Table \ref{compare}. The \textit {glasso} without adjusting
for the effects of genetic markers resulted in a total of 18,987 edges
with an optimal tuning parameter $\lambda=0.22$. There were 9,854 links
that were selected by both procedures. Again \textit{glasso} selected a~lot more links than the \textit{cGGM}; among the links that were
identified by the \textit{glasso} only, 1,569 pairs are associated with at
least one common genetic marker (see Table \ref{compare}), further
explaining that some of the links identified by gene expression data
alone can be due to shared comment genetic variants.

The modified neighborhood selection procedure \textit{mLasso}
identified only 1,917 edges with $\lambda=0.42$, out of which 1,750
were identified by the \textit{cGGM} and 1,916 were identified by the
\textit{glasso}. There was a common set of 1,749 links that were identified
by all three procedures. A summary of the degrees of the graphs
estimated by these three procedures is given in Table \ref{degree}.
We observe that the \textit{glasso} gave a much denser graph than the other
two procedures, agreeing with what we observed in simulation
studies.

\begin{figure}
\begin{tabular}{@{}cc@{}}

\includegraphics{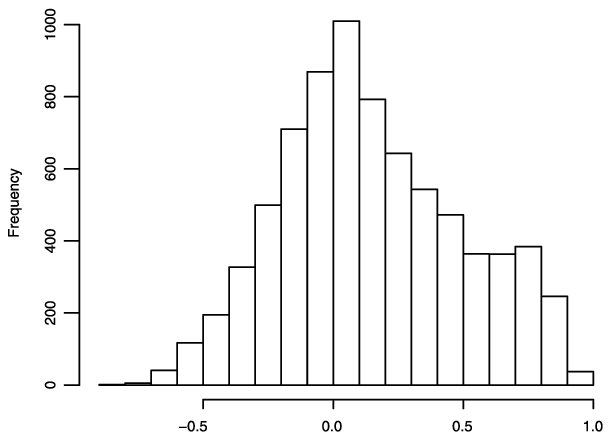}
 & \includegraphics{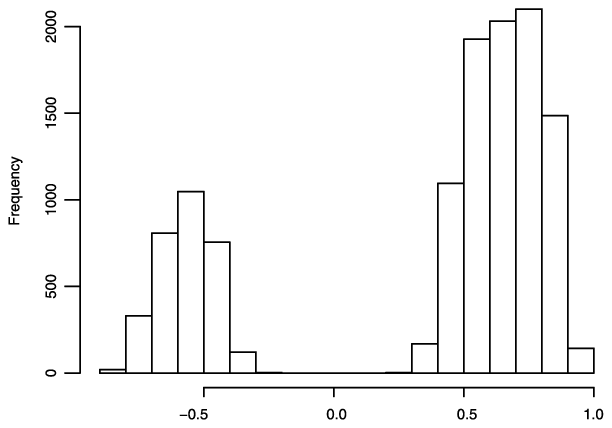}\\
(a) & (b)\\[4pt]

\includegraphics{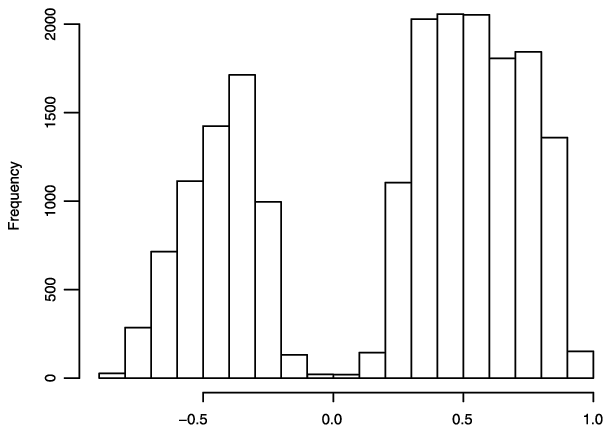}
 & \includegraphics{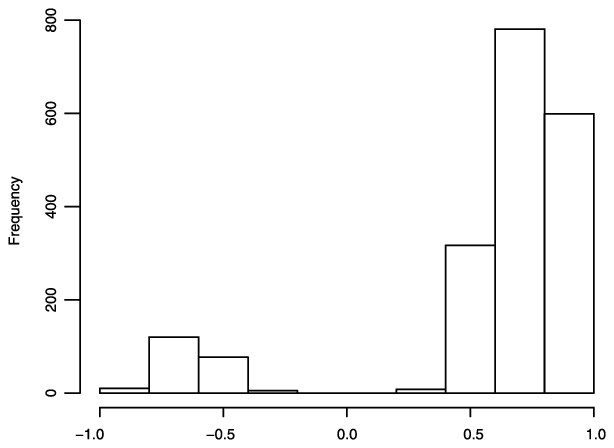}\\
(c) & (d)
\end{tabular}
\caption{Histograms of marginal correlations for pairs of linked genes
based on BioGRID \textup{(a)} and linked genes identified by
\textit{cGGM} \textup{(b)}, \textit{glasso} \textup{(c)} and a modified
neighborhood selection procedure (\textit{mLasso}) \textup{(d)}.}
\label{hist}
\end{figure}

If we treat the PPI of the BioGRID database as the true network
among these genes, the true positive rates from \textit{cGGM}, \textit
{glasso} and the modified neighborhood selection procedure were 0.067,
0.071 and 0.019, respectively, and the false positive rates were
0.016, 0.026 and 0.0025, respectively. The MCC scores from \textit
{cGGM}, \textit{glasso} and the modified neighborhood selection
procedure were 0.041, 0.030 and 0.033, respectively. One reason for
having low true positive rates is that many of the protein--protein
interactions cannot be reflected at the gene expression level.
Figure \ref{hist}(a) shows the histogram of the correlations of
%
%
genes that are linked on the BioGRID PPI network, indicating that
many linked gene pairs have very small marginal correlations. The
Gaussian graphical models are not able to recover these links.
Figure~\ref{hist} plots (b)--(d) show the marginal correlations of
the genes pairs that were identified by \textit{cGGM}, \textit{glasso} and
\textit{mGlasso}, clearly indicating that the linked genes identified
by the \textit{cGGM} have higher marginal correlations. In contrast,
some linked genes identified by \textit{glasso} have quite small
marginal correlations. Another reason is that the PPI represents the
marginal pair-wise interactions among the proteins rather than the
conditional interactions.\looseness=1

\section{Conclusions and discussion}
We have presented a sparse conditional Gaussian graphical model for
estimating the sparse gene expression network based on eQTL data in
order to account for genetic effects on gene expressions. Since
genetic variants are associated with expression levels of many
genes, it is important to consider such heterogeneity in estimating
the gene expression networks using the Gaussian graphical models. We
have demonstrated by simulation studies that the proposed sparse
\textit{cGGM} can estimate the underlying gene expression networks more
accurately than the standard \textit{GGM}. For the yeast eQTL data set
we analyzed, the standard Gaussian graphical model without adjusting
for possible genetic effects on gene expressions identified many
possible false links that result in very dense graphs and make the
interpretation of the resulting networks difficult. On the other
hand, our proposed \textit{cGGM} resulted in a much sparser and
biologically more interpretable network. We expect similarly good
performance on data from other published sources, such as from
\citet{Schetal03} and \citet{CheSpi02}.

Due to the limits of the gene expression data, one should not expect
to recover completely the true signaling networks since many
dependencies among these genes can be observed only at the protein
or metabolite level. In any global biochemical network such
signaling network or protein interaction network, genes do not
interact directly with other genes; instead, gene induction or
repression occurs through the activation of certain proteins, which
are products of certain genes [\citet{BraFueMen02}, \citet{Kon}]. Similarly, gene transcription can also be affected by
protein-metabolite complexes. Despite these limitations of the gene
expression, it is still useful to abstract the actions of proteins
and metabolites and represent genes acting on other genes in a gene
network [\citet{Kon}]. This gene network is what we aim to learn
based on the proposed \textit{cGGM}. As we observed from our analysis
of the yeast eQTL data, such graphs or gene networks constructed
from the \textit{cGGM} can indeed explain the data and provide certain
biological insights into gene interactions. Such graphs can be
interpreted as a projection of true signaling or protein interaction
network into the gene space [\citet{BraFueMen02}, \citet{Kon}].

We have focused in this paper on estimating the sparse conditional
Gaussian graphical model for gene expression data by adjusting for the
genetic effects on gene expressions. However, we expect that by
explicitly modeling the covariance structure among the gene
expressions, we should also improve the identification of the genetic
variants associated with the gene expressions
[\citet{RotLevZhu10}]. This is in fact the original motivation of
the SUR models proposed by \citet{Zel62}. It would be interesting
to investigate theoretically as to how modeling the concentration
matrix can lead to improvement in estimation and identification of the
genetic variants associated with the gene expression traits.

We used the Gaussian graphical models for studying the conditional
independence among genes at the transcriptional level. Such
undirected graphs do not provide information on causal dependency.
Data from genetic genomics experiments have been proposed to
construct the gene networks represented by directed causal graphs.
For example, \citet{LiuDeLHoe08} and \citet{BinHoe05}
used structural equation modeling and a genetic algorithm to
construct causal genetic networks among genetic loci and gene
expressions. \citet{CNetal10} developed an efficient Markov
chain Monte Carlo algorithm for joint inference of causal network
and genetic architecture for correlated phenotypes. Although
genetical genomics data can indeed provide opportunity for inferring
the causal networks at the transcriptional level, these causal
graphical model-based approaches can often only handle a small number
of transcripts because the number of possible directed graphs is
super-exponential in the number of genes considered
[\citet{ChiHecMee03}]. Regularization methods may provide alternative
approaches to joint modeling of genetic effects on gene expressions
and causal graphs among genes at the expression level.

\section*{Acknowledgments}

We thank the three reviewers and the Editor for many insightful
comments that have greatly improved the presentation of this paper.

\begin{supplement}[id=suppA]
\stitle{Supplemental materials for ``A sparse conditional Gaussian
graphical model for analysis of
genetical genomics data''\\}
\slink[doi,text={10.1214/11-\break
AOAS494SUPP}]{10.1214/11-AOAS494SUPP} 
\slink[url]{http://lib.stat.cmu.edu/aoas/494/supplement.pdf}
\sdatatype{.pdf}
\sdescription{The online supplemental materials
include the simulation standard errors of Tables \ref{simu.tb2}
and \ref{simu.tbl}, two
propositions on the Hessian matrix of the likelihood function and
the convergence of the algorithm and the theoretical properties of
the proposed penalized estimates of the sparse \textit{cGGM}: its
asymptotic distribution, the oracle properties when~$p$ and $q$ are
fixed as $n\rightarrow\infty$ and the convergence rates and
sparsistency of the estimators when $p=p_n$ and $q=q_n$ diverge as
$n\rightarrow\infty$. All the proofs are also given in the
supplemental materials.}
\end{supplement}

%

\printaddresses

\end{document}